\documentclass[11pt,twoside]{article}
\usepackage{./asp2014}

\aspSuppressVolSlug
\resetcounters

\begin{document}

\title{Clumpy molecular structures revolving the \protect{B[e]} supergiant MWC\,137}
\author{M. Kraus,$^{1,2}$ L.S. Cidale,$^{3,4}$ T. Liimets,$^{2,5}$ C.E. Cappa,$^{6}$ N. Duronea,$^{6}$\\ D.S. Gunawan,$^{7}$ M.E. Oksala,$^{8}$ M. Santander-Garc\'{i}a,$^{9}$ M.L. Arias,$^{3,4}$\\ D.H. Nickeler,$^{1}$ G. Maravelias,$^{1}$ M. Borges Fernandes,$^{10}$ and M. Cur\'{e}$^{7}$
\affil{$^1$Astronomick\'{y} \'{u}stav AV\,\v{C}R, v.v.i., Ond\v{r}ejov, Czech Republic; \email{michaela.kraus@asu.cas.cz}\\
$^2$Tartu Observatory, T\~oravere, Estonia\\
$^3$Facultad de Ciencias Astron\'{o}micas y Geof\'{i}sicas, UNLP, La Plata, Argentina\\
$^4$Instituto de Astrof\'{i}sica de La Plata, La Plata, Argentina\\
$^5$Institute of Physics, University of Tartu, Tartu Estonia\\
$^6$Instituto Argentino de Radioastronom\'{i}a, La Plata, Argentina\\
$^7$Universidad de Valpara\'{i}so, Valpara\'{i}so, Chile\\
$^8$LESIA, Observatoire de Paris, Meudon, France\\
$^9$Instituto de Ciencia de Materiales de Madrid (CSIC), Madrid, Spain\\
$^{10}$Observat\'{o}rio Nacional, Rio de Janeiro, Brazil}}

% This section is for ADS Processing.  There must be one line per author.
\paperauthor{Michaela~Kraus}{michaela.kraus@asu.cas.cz}{}{Astronomick\'y \'ustav, Akademie v\v{e}d \v{C}esk\'e republiky, v.v.i.}{}{Ond\v{r}ejov}{}{25165}{Czech Republic}
\paperauthor{Lydia~Cidale}{lydia@fcaglp.unlp.edu.ar}{}{Facultad de Ciencias Astron\'omicas y Geof\'\i sicas (UNLP) and IALP (CONICET-UNLP)}{}{La Plata}{Buenos Aires}{1900}{Argentina}
\paperauthor{Tiina~Liimets}{tiina@obs.ee}{}{Tartu Observatory}{}{T\~oravere}{Tartumaa}{61602}{Estonia}
\paperauthor{Cristina~Cappa}{cristina.elisabet.cappa@gmail.com}{}{Instituto Argentino de Radioastronom\'{i}a}{}{La Plata}{}{B1900FWA}{Argentina}
\paperauthor{Nicolas~Duronea}{duronea@iar.unlp.edu.ar}{}{Instituto Argentino de Radioastronom\'{i}a}{}{La Plata}{}{B1900FWA}{Argentina}
\paperauthor{Diah~Gunawan}{diah.gunawan@uv.cl}{}{Instituto de F\'{i}sica y Astronom\'{i}a}{Facultad de Ciencias, Universidad de Valpara\'{i}so}{Valpara\'{i}so}{}{5030}{Chile}
\paperauthor{Mary~Oksala}{meo@udel.edu}{}{California Lutheran University}{Department of Physics}{Thousand Oaks}{California}{91360}{USA}
\paperauthor{Miguel~Santander-Garc\'{i}a}{msantander@icmm.csic.es}{}{Instituto de Ciencia de Materiales de Madrid (CSIC)}{}{Madrid}{}{28049}{Spain}
\paperauthor{Maria~Laura~Arias}{mlaura@fcaglp.fcaglp.unlp.edu.ar}{}{Facultad de Ciencias Astron\'omicas y Geof\'\i sicas (UNLP) and IALP (CONICET-UNLP)}{}{La Plata}{Buenos Aires}{1900}{Argentina}
\paperauthor{Dieter~Nickeler}{dieter.nickeler@asu.cas.cz}{}{Astronomick\'y \'ustav, Akademie v\v{e}d \v{C}esk\'e republiky, v.v.i.}{}{Ond\v{r}ejov}{}{25165}{Czech Republic}
\paperauthor{Grigoris~Maravelias}{grigorios.maravelias@asu.cas.cz}{}{Astronomick\'y \'ustav, Akademie v\v{e}d \v{C}esk\'e republiky, v.v.i.}{}{Ond\v{r}ejov}{}{25165}{Czech Republic}
\paperauthor{Marcelo~Borges~Fernandes}{borges@on.br}{}{Observat\'orio Nacional}{}{S\~ao Cristov\~ao, Rio de Janeiro}{}{20921-400}{Brazil}
\paperauthor{Michel~Cur\'{e}}{michel.cure@uv.cl}{}{Instituto de F\'{i}sica y Astronom\'{i}a}{Facultad de Ciencias, Universidad de Valpara\'{i}so}{Valpara\'{i}so}{}{5030}{Chile}

%\paperauthor{Sample~Author2}{Author2Email@email.edu}{ORCID_Or_Blank}{Author2 Institution}{Author2 Department}{City}{State/Provincyye}{Postal Code}{Country}

\begin{abstract}
The peculiar emission-line star MWC\,137 with its extended optical nebula was 
recently classified as B[e] supergiant. To study the spatial distribution of 
its circumstellar molecular gas on small and large scales, we obtained 
near-infrared and radio observations using SINFONI and APEX, respectively. We 
find that the hot CO gas is arranged in moving clumpy ring and shell structures 
close to the star, while a cold CO envelope is encircling the borders of the 
optical nebula from the south to the west.
\end{abstract}

\section{Introduction}

The Galactic object MWC\,137 is a peculiar early-type star surrounded
by the optical nebula Sh\,2-266 (80$'' \times$ 60$''$) of unclear origin.
A large-scale collimated outflow with several knots was recently detected 
in the light of the [N\,{\sc ii}]\,6583 line \citep{2016A&A...585A..81M}. 
Moreover, near-infrared spectroscopic observations displayed intense, 
kinematically broadened CO band emission in both isotopes $^{12}$CO and 
$^{13}$CO \citep{2013A&A...558A..17O}. The observed enrichment in $^{13}$CO 
implies that MWC\,137 is an evolved object \citep{2015AJ....149...13M},
and \citet{2016A&A...585A..81M} confirmed its supergiant nature.

\section{Observations and Results}

We obtained SINFONI $K$-band IFU spectroscopic data of MWC\,137 on 2014 
December 30 and 2016 March 19 with high-spatial resolution (FOV of $0.8\arcsec
\times 0.8\arcsec$). The continuum subtracted hot CO band images (Fig.\,1, 
left) display an outer ring (shell?) with $r_{\rm out} = 225$\,mas (dashed 
circle) and an inner disk or ring (ellipse) with two large blobs (pointed at by 
the arrows). The major and minor semiaxes are 
112.5\,mas and 97.5\,mas, resulting in an inclination of $\sim 30^{\circ}$.
These were determined by the position of the maximum intensity of the blobs and
the constraint that the disk should be roughly perpendicular to the optical jet.
The two blobs show an angular motion of $\sim 10\deg$ within 15
months. This would translate into $\varv_{\rm rot} = 375$\,km\,s$^{-1}$, if we
assume a distance of 5.2\,kpc, which is too fast for Keplerian rotation.

Observations of the $^{12}$CO(3-2) line at 345\,GHz were obtained with the 
Atacama Pathfinder EXperiment (APEX) in a region of $3\arcmin\times 3\arcmin$ 
centered on Sh2-266, with an angular resolution of $20\arcsec$.
The cold CO emission comprises a partial shell in the velocity interval 
[+27.3,+30.3]\,km\,s$^{-1}$ (contours in Fig.\,1, right). 
According to circular galactic rotation models and the velocity field of the 
Galaxy by \citet{1993A&A...275...67B}, gas at these velocities is located at 
kinematical distances $d=5-9$\,kpc, in good agreement with the estimates of
\citet{2016A&A...585A..81M}.

\articlefigure[width=0.9\textwidth]{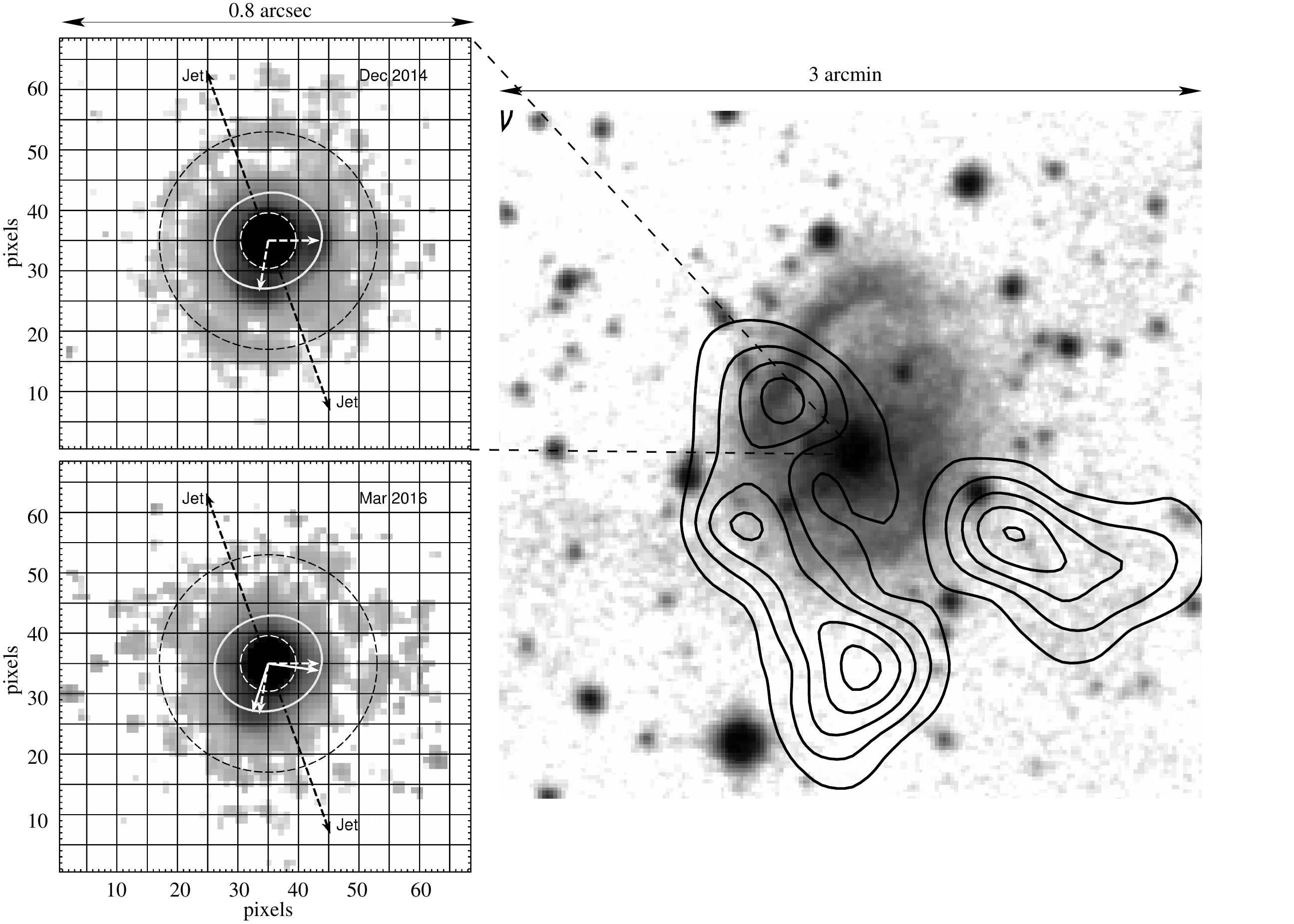}{fig}{Location and variation of 
the hot, small-scale (left) and the cold, large-scale (right, contours) CO
emission with respect to the jet and the optical nebula.}

\acknowledgements Observations were obtained under ESO programs 094.D-0637
and 097.D-0033. We acknowledge support from GA\,\v{C}R (14-21373S), 
RVO:67985815, 
and ETAg (IUT40-1). 
%and RVO:67985815. 

\bibliography{Kraus_bep2016_poster}  % For BibTex

% For non-BibTex:

\end{document}